\documentclass[twoside]{dis08}
\usepackage[latin1]{inputenc}
\usepackage[dvips]{graphicx,epsfig,color}
\usepackage{wrapfig,rotating}
\usepackage{amssymb,amsmath,array}

\begin{document}
\title{Charm Physics at CDF}

%***********************************************************************
% AUTHORS INFORMATION AREA
%***********************************************************************
\author{P. J. Bussey 
%
% Optional short acknowledgment: remove next line if non-needed
\thanks{Royal Society of Edinburgh Scottish Executive\newline
Support Research Fellow.}
%
% DO NOT MODIFY THE FOLLOWING '\vspace' ARGUMENT
\vspace{.3cm}\\
%
% Addresses and institutions (remove "1- " in case of a single institution)
Department of Physics and Astronomy, \\
University of Glasgow, Glasgow G12 8QQ, U.K.
%
% Remove the next three lines in case of a single institution
\vspace{.1cm}\\
for the CDF Collaboration. \\
}
%***********************************************************************
% END OF AUTHORS INFORMATION AREA
%***********************************************************************

\maketitle

\begin{abstract}
A survey of recent results in charm physics from CDF is presented.
\end{abstract}

\section{Introduction}

Since the start of the Tevatron Run II running at 1.96 TeV in 2001,
the CDF experiment has accumulated extremely large samples of
charm-containing events.  These have been taken with several triggers,
and especially the so-called ``two track trigger'' which selects
tracks displaced from the main vertex at an early stage of the
trigger process.  Other useful triggers have been based on muon
identification. In this talk I present three recent analyses which
illustrate different facets of the CDF charm program.

\section{Inclusive $\psi(2s)$ production}

\begin{wrapfigure}{r}{0.5\columnwidth}
\centerline{\includegraphics[width=0.45\columnwidth]{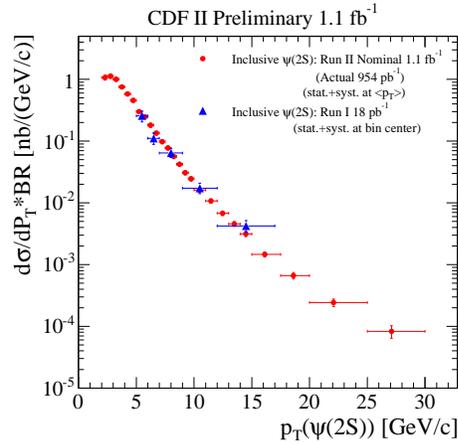}}
\caption{Inclusive cross sections for $\psi'$ production 
compared with the Run I results }\label{Fig:1}
\end{wrapfigure}

In Run I, CDF published inclusive cross sections for both $J/\psi$ and
$\psi'$ production~\cite{CDF1}, and an earlier CDF II
paper~\cite{CDFjpsi} presented inclusive cross sections for $J/\psi$
production.  These results have now been complemented 
with a new measurement of the inclusive $\psi'$ cross section using
1.1 fb$^{-1}$ of Run II data.  The central drift chamber and
muon detection system were used to detect the dimuon decay of the $\psi'$,
selecting on dimuon masses in the range 3.5 - 3.8 GeV/c$^2$, with 
the $\psi'$ rapidity in the range $\pm0.6$, and transverse
momentum in the range 2.0 - 30.0  GeV/c, giving a clear improvement on the 
Run 1 lower limit of 5.0 GeV/c.

The signal was extracted from the background by means of an unbinned
maximum likelihood fit using parametrized functions for the background
and the signal.  The signal and background both have a prompt
$\psi'$ component and one arising from $B$ decays, which are
distinguished by means of the distance of the dimuon vertex from the
beamline as measured using the Silicon Vertex Detector.  The prompt
component was modelled as a Gaussian in terms of the distance of the
vertex from the beamline, and the decayed component by convoluting this with
an exponential decay distribution.

\begin{figure}[t!]
~\\[-0ex]
\centerline{\includegraphics[width=0.45\columnwidth]{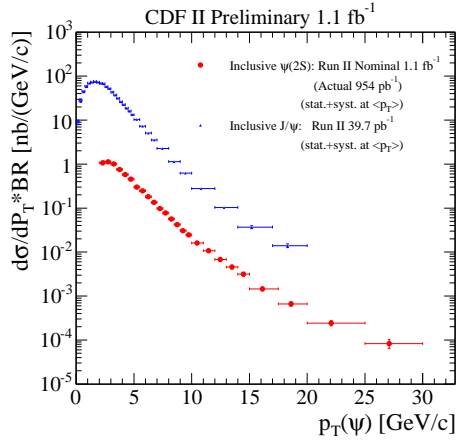}}
\caption{Inclusive cross sections for $\psi'$ production at CDF,
as in Fig.\ 1, compared to the $J/\psi$ cross section.
}\label{Fig:2}
\end{figure}
The results are shown in Figs.\ 1 - 3.  The total integrated  $\psi'$
cross section is $3.141\pm0.038$(stat)$^{+0.225}_{-0.218}$(sys) nb. There is good
agreement with the Run I results.  In comparison with the 
$J/\psi$ cross section, that of the $\psi'$ is lower by one to
two orders of magnitude.
\begin{figure}[h!]
\centerline{
\includegraphics[width=0.45\columnwidth]{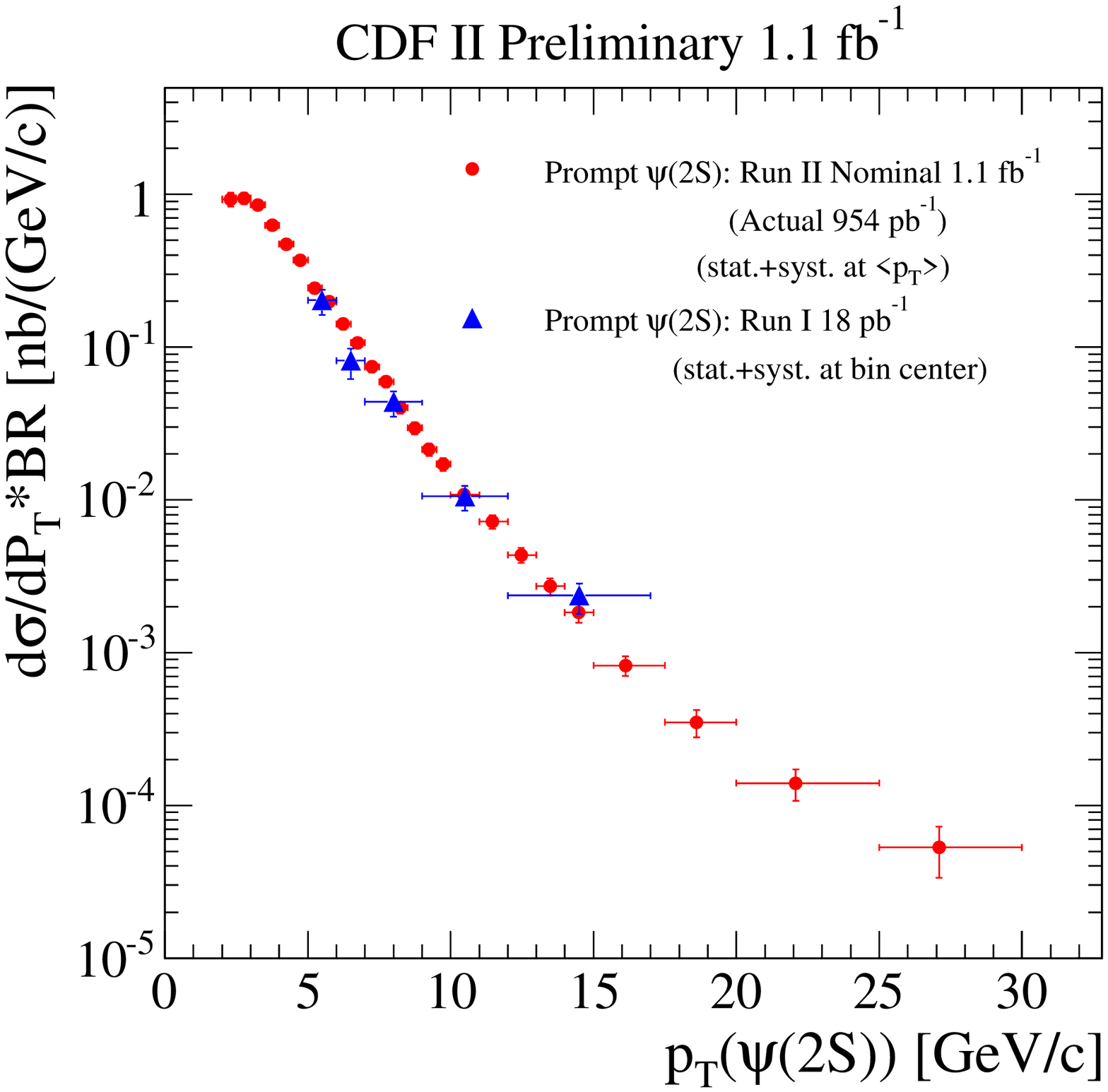}\\
\includegraphics[width=0.45\columnwidth]{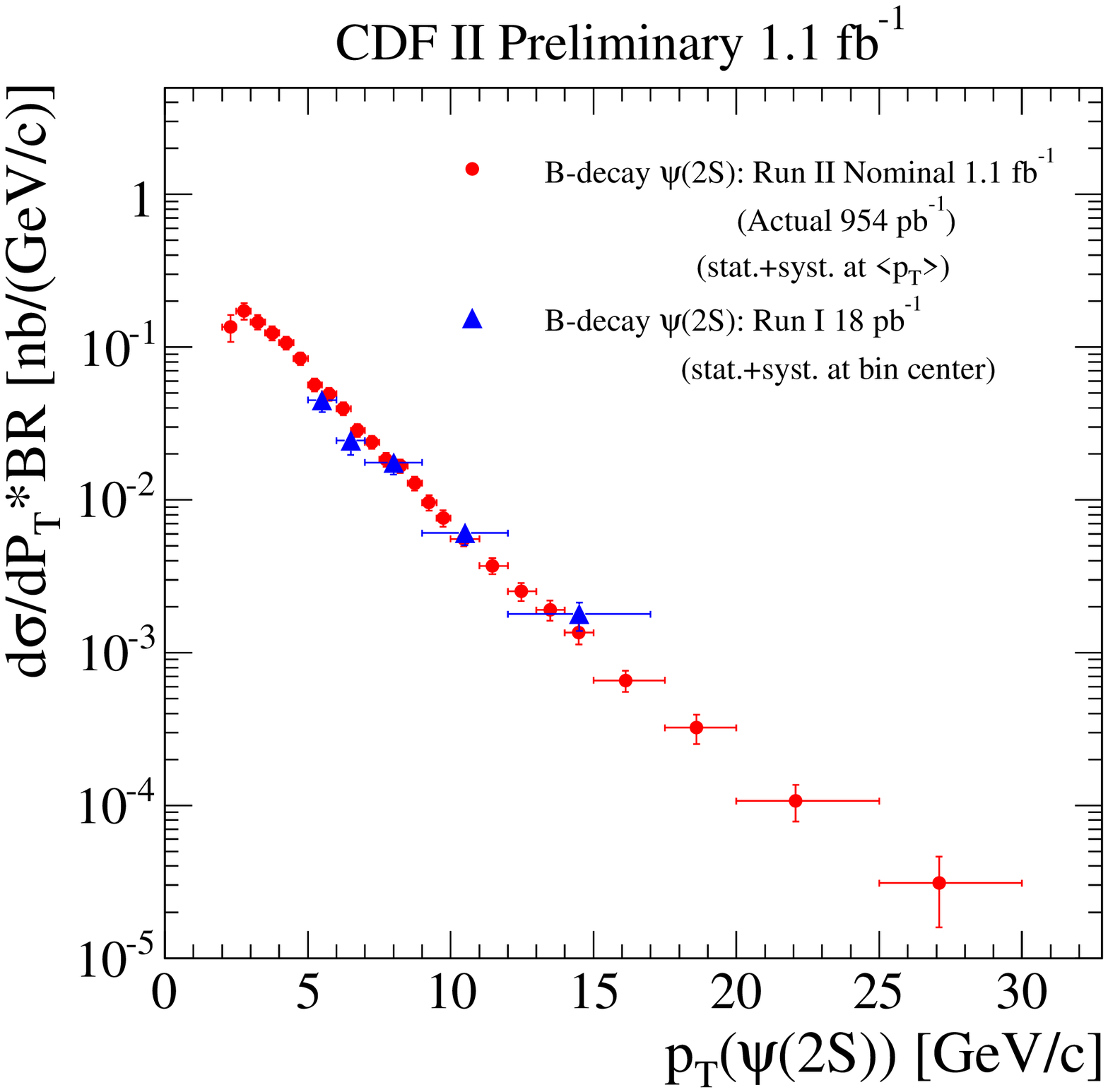}
}
\caption{Inclusive cross sections for $\psi'$ production at CDF,
showing the prompt (upper) and $B$-decay (lower) components separately, compared
to the Run I results.
}\label{Fig:3}
\end{figure}

\section{$D^0$ mixing at the Tevatron}

There have recently been significant developments in extending the
studies of mixing into the charm sector. BELLE have obtained evidence
that the lifetime for $D^0$ decay into CP eigenstates, such as
$K^+K^-$ and $\pi^+\pi^-$ is different from CP-mixed states, such as
$K^+\pi^-$ \cite{Belle}.  BaBar, meanwhile, have found a difference in
the decay time between the decays $D^0\to K^+\pi^-$ and $D^0\to
K^-\pi^+$.  CDF have now found evidence for the latter of these two
effects, for the first time in $D$ mesons produced by hadronic
collisions.
\cite{DDmix}.

The approach followed is first to define the quantity
$R(t)$ as the ratio of the observed numbers of $K^+\pi^-$ and
$K^-\pi^+$ decays undergone by $D^0$ under experimental
conditions. The similarity of the kinematic acceptances means that the
systematic uncertainty on this measured ratio is small.  Then, with 
reasonable assumptions, it can be shown that 
$$R(t) = R_D + \sqrt{R_D}y't + \frac{1}{4}(x'^2 + y'^2)t^2,$$
where $R_D$ is the ``direct'' ratio, $t$ is the decay time and $x',\,y'$ are
linear combinations of 
$x = \Delta m/\Gamma$ and $y = \Delta\Gamma/2\Gamma$, namely the scaled
mass and width differences between the two physical eigenstates.

\begin{figure}[t!]
~\\[-4ex]
\centerline{
\raisebox{5ex}{\includegraphics[width=0.37\columnwidth]{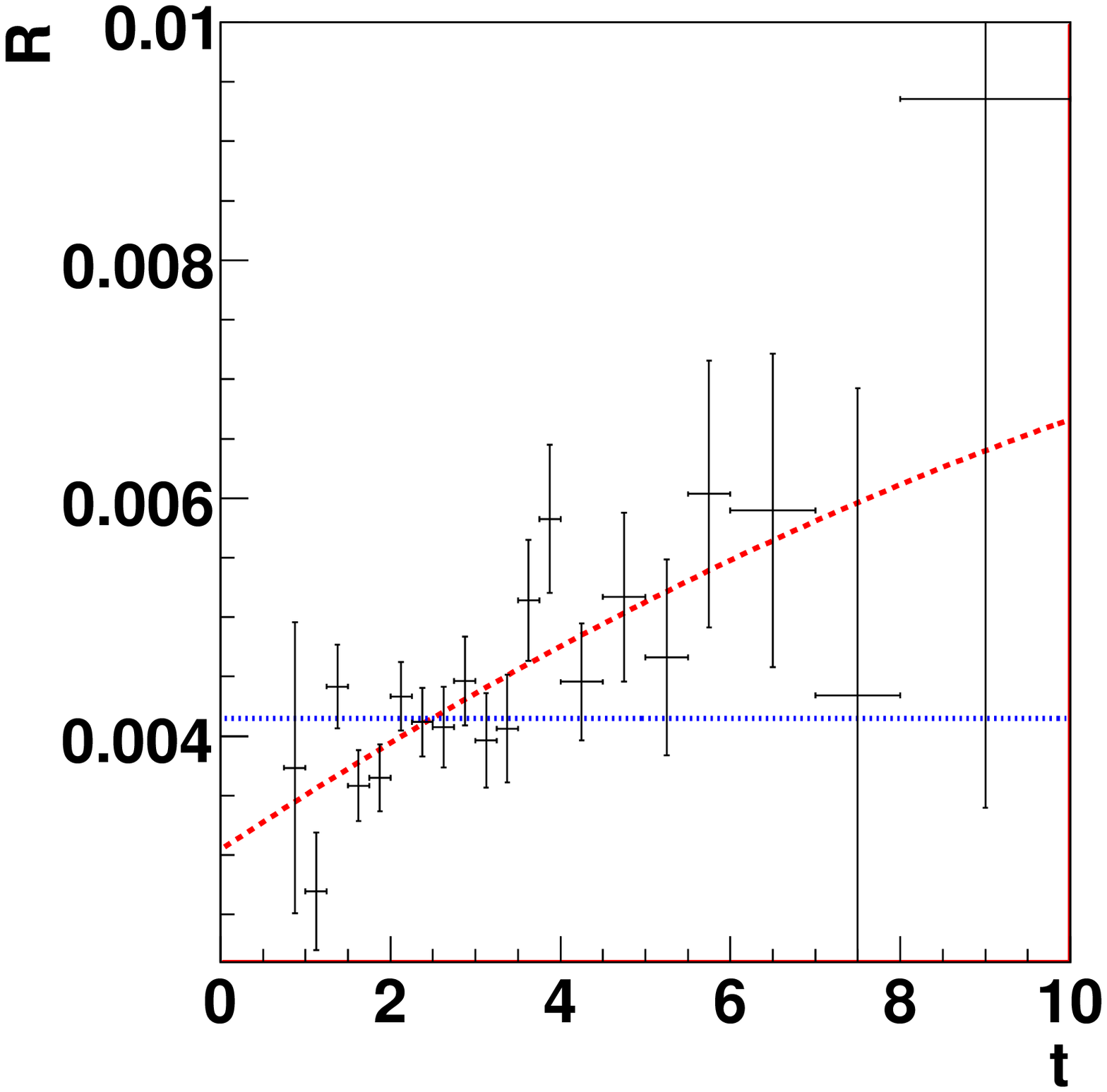}}
\raisebox{0ex}{\includegraphics[width=0.45\columnwidth]{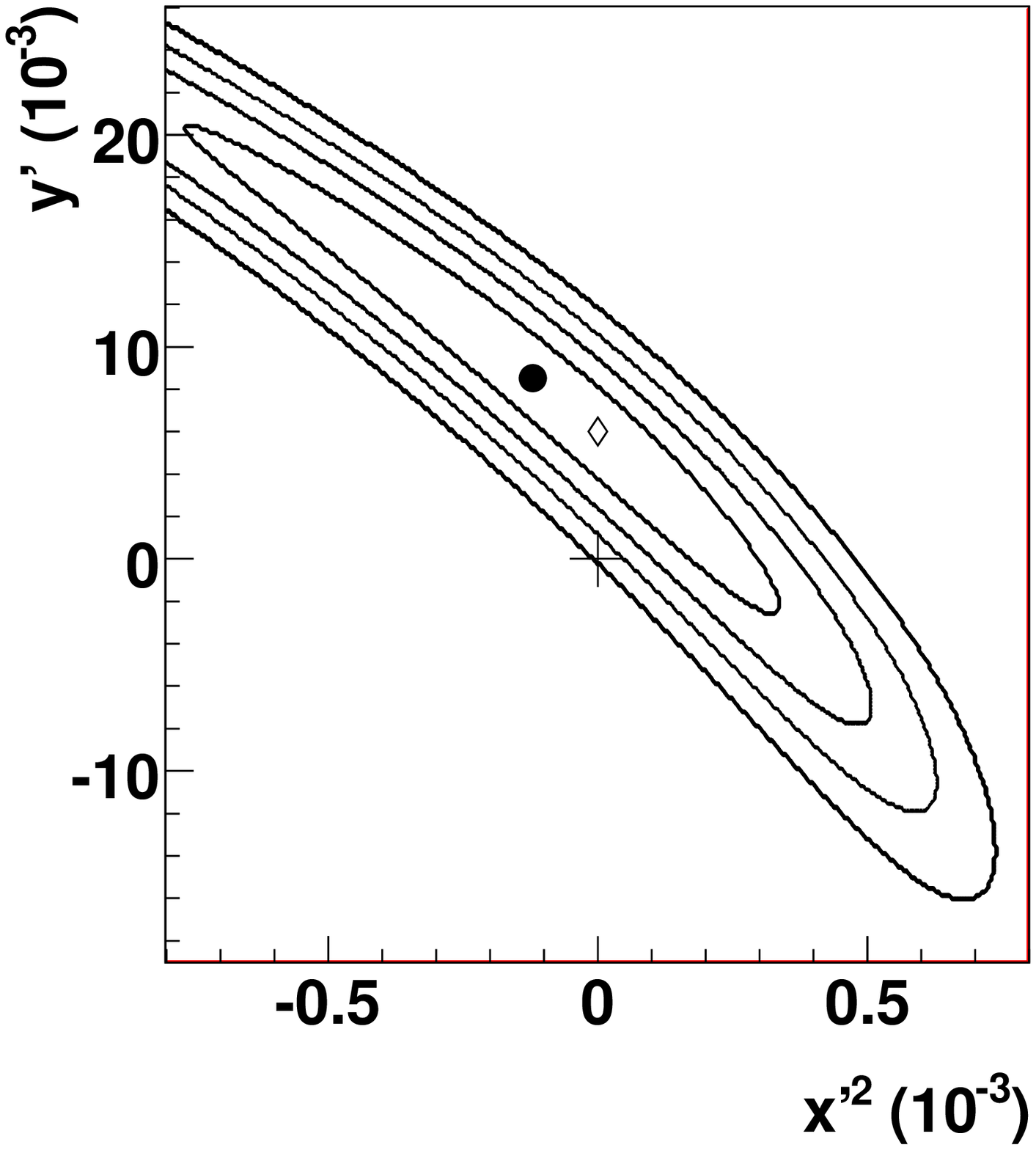}}}
~\\[-10ex]
\caption{Left: ``wrong / right sign'' ratio for $D^0$ decays,
plotted against proper decay time in units of the mean decay time,
with a fitted quadratic curve.  Right: probability contours in $x',\,
y'$, corresponding to 1 - 4 Gaussian standard deviations. The closed
circle and open diamond, show central results from an unconstrained and a
physically allowed fit. The cross shows the no-mixing point.  The
parameter $R_D$ is removed by a Bayesian integration.  }\label{Fig:5}
\end{figure}

Using 1.5 fb$^{-1}$ of data, the initial $D^0$ or $\overline{D}^0$
mesons are identified in $D^*$ decays, using the sign of the soft pion
to tag the $D$ state. The ratio $R(t)$ of the``wrong sign / right
sign'' decays (Cabibbo doubly suppressed / allowed, in the
absence of oscillations) is then plotted as a function of decay time.
The result, when fitted with a quadratic curve, is suggestive that
$R(t)$ rises with $t$ (Fig.\ 4). Bayesian probability contours were
now constructed in the $x', y'$ space using a flat prior. A value of
$x', y'$ significantly different from the no-mixing point at the
origin indicates $D^0$-$\overline{D}^0$ mixing.  The no-mixing point
is found to lie on a contour corresponding to a probability level of
$1.5\times10^{-4}$ or 3.8 $\sigma$. This is claimed as evidence for
mixing.

\section{Search for the rare decay $D^0\to\mu^+\mu^-$}
The decay $D^0\to\mu^+\mu^-$ is strongly suppressed in the Standard Model.
Figure 5 shows two short-range SM decays (top) a long-range SM
decay (middle) and a possible SUSY decay (bottom). The
predicted SM rates are very low and correspond to a branching
ratio of approximately $4\times10^{-13}$, far below 
foreseeable experimental measurements.  However the SUSY
decays can give branching ratios of up
to $3.5\times10^{-6}$, well within present experimental capabilities.
Earlier in Run II, CDF published a branching ratio limit of
$2.5\times10^{-6}$ (90\% CL) \cite{CDFmumu1}, which has since been
reduced to $1.3\times10^{-6}$ by BaBar \cite{mumubabar}. Here using a
larger data sample, we present a new result which improves
substantially on the previous values.

\begin{figure}[t!]
\parbox{0.45\columnwidth}{
\hspace*{1ex}\includegraphics[width=0.45\columnwidth]{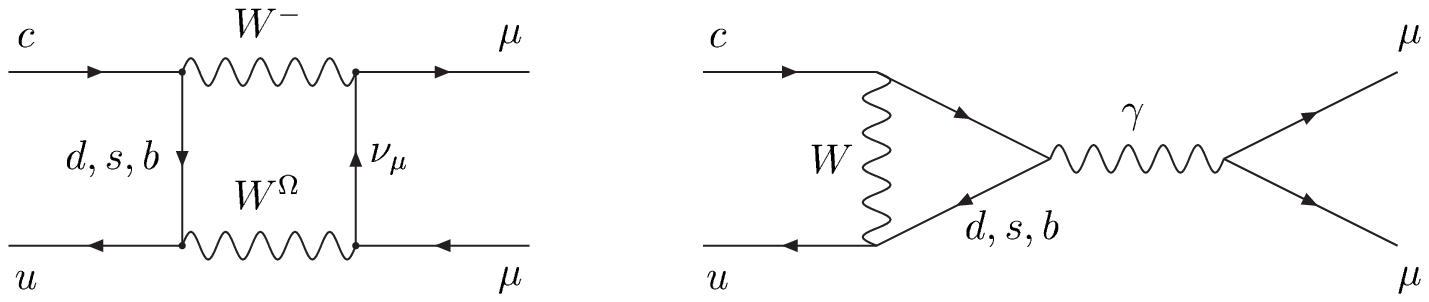}\protect\\
\hspace*{11ex}\includegraphics[width=0.25\columnwidth]{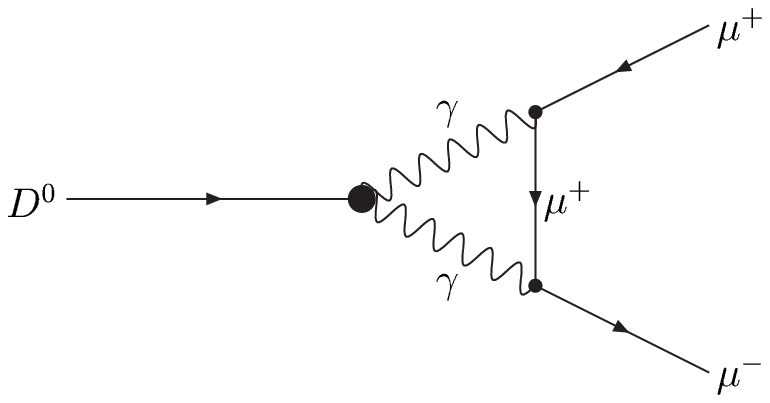}\\
\hspace*{11ex}\includegraphics[width=0.2\columnwidth]{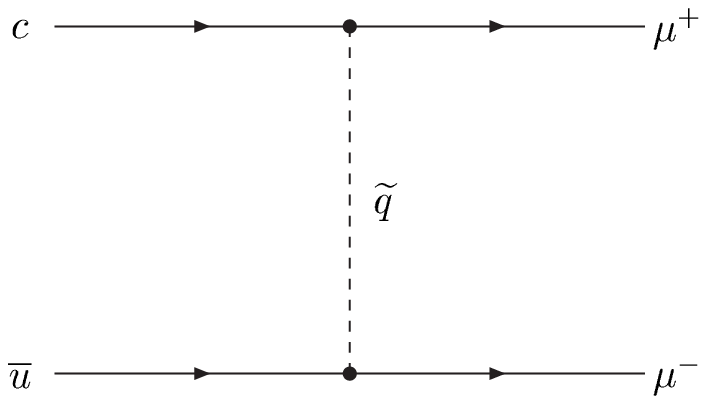}
}
\hspace*{0.1\columnwidth}
\parbox{0.45\columnwidth}{
\includegraphics[width=0.4\columnwidth]{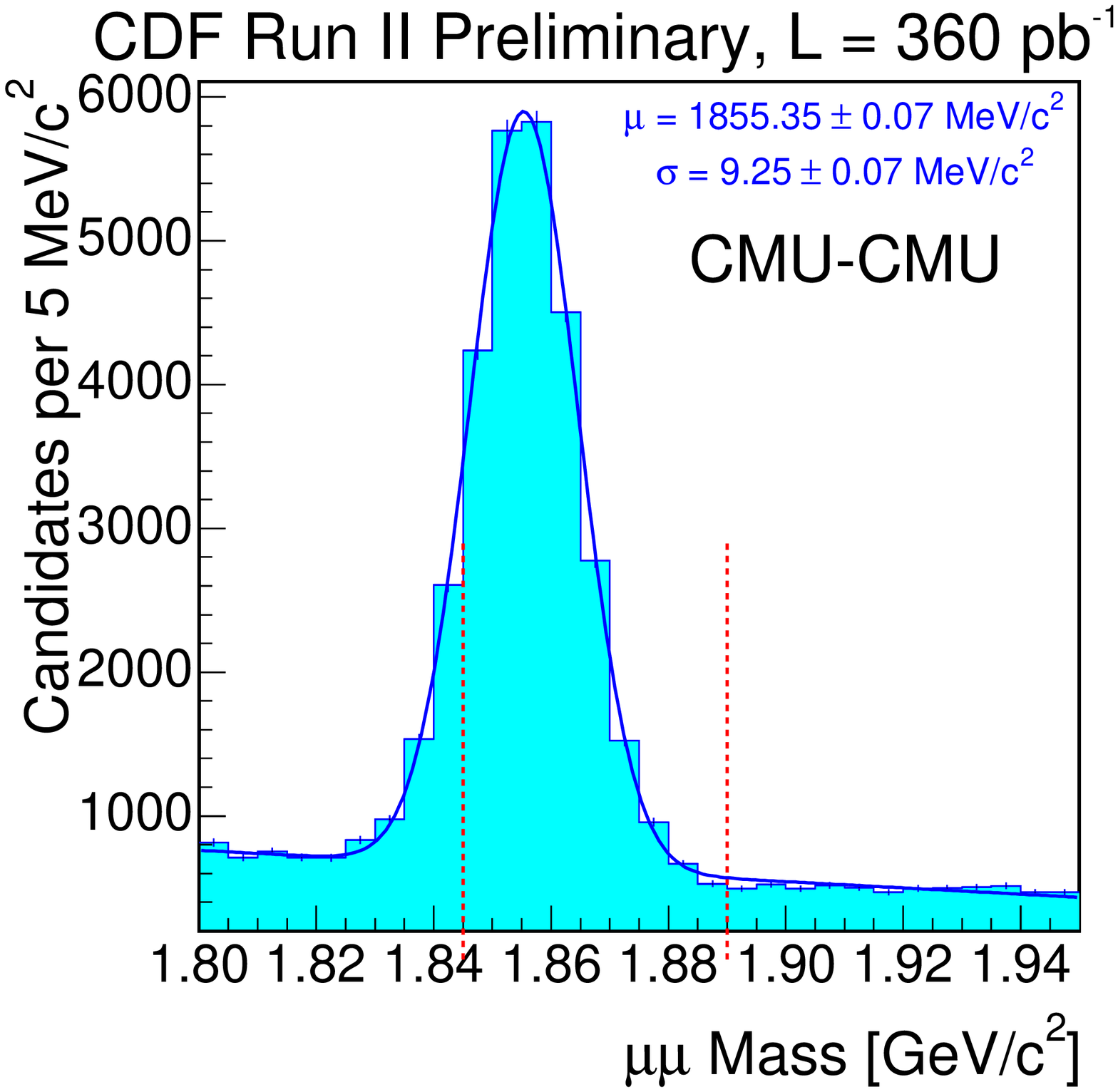}}
\\Figure 5: Feynman diagrams for the decay \hfill Figure 6: Reference peak for the $\pi^+\pi^-$ decay,\protect\\
 $D^0\to\mu^+\mu^-$. \hfill indicating
the $D\to\mu^+\mu^-$ mass region.
\label{Fig:6}
\end{figure}

In order to reduce backgrounds, which arise principally from muon
pairs from $B$ meson decays, both decay muons are required to have
good identification within the CDF central muon detector system (CMU,
CMX). An illustration of the mass peak obtainable in CDF is given in
Fig.\ 6, which shows $\pi^+\pi^-$ decays of $D^0$ meson in which the
pions have been relabelled as muons and point into the central muon
detectors.  The muons arising from $B$ decays are removed by cuts on
the vertex position and pointing of the muon pair as measured in the
Silicon Vertex Detector.  A total of 4 candidate events were observed,
with an expectation of 8.6; using a Bayesian approach this gives an
upper limit on the branching ratio of $4.3\times10^{-7}$ (90\% CL) or
$5.3\times10^{-7}$ (95\% CL). Using just 360 pb$^{-1}$ of data, this
is currently the world's best result for this channel, and translates
into significant constraints on the R-parity violating couplings
within SUSY.

% ****************************************************************************
% BIBLIOGRAPHY AREA
% ****************************************************************************

\begin{footnotesize}
% IF YOU DO NOT USE BIBTEX, USE THE FOLLOWING SAMPLE SCHEME FOR THE REFERENCES
% ----------------------------------------------------------------------------

% ----------------------------------------------------------------------------

% IF YOU USE BIBTEX,
% - DELETE THE TEXT BETWEEN THE TWO ABOVE DASHED LINES
% - UNCOMMENT THE NEXT TWO LINES AND REPLACE 'Name_Of_Your_BibFile'

%\bibliographystyle{unsrt}
%\bibliography{Name_Of_Your_BibFile}
% example of Name_Of_Your_BibFile.bib
% @Article{Turcato:2006ch,
%      author    = "Turcato, M.",
%  collaboration = "ZEUS and H1",
%      title     = "Lepton flavour violation and charmonium physics at HERA",
%      journal   = "Nucl. Phys. Proc. Suppl.",
%      volume    = "162",
%      year      = "2006", 
%      pages     = "283-287",
%      SLACcitation  = "%%CITATION = NUPHZ,162,283;%%"
% }
% 
% @Unpublished{Gogitidze:2007du,
%      author    = "Gogitidze, N.",
%  collaboration = "H1", 
%      title     = "Prompt photons and particle momentum distributions at
%                   HERA", 
%      year      = "2007",
%      note    = "hep-ex/0701033",
%      SLACcitation  = "%%CITATION = HEP-EX 0701033;%%"
% }

\end{footnotesize}

% ****************************************************************************
% END OF BIBLIOGRAPHY AREA
% ****************************************************************************

\end{document}